\documentclass[a4paper]{article}

\usepackage{INTERSPEECH2020}
\usepackage{graphicx}
\usepackage{tabularx}
\usepackage{comment}
\usepackage{xcolor}
\usepackage{subcaption}
\usepackage[flushleft]{threeparttable} 
\usepackage[redeflists]{IEEEtrantools}

\renewcommand{\textcolor}[2]{#2}

\title{Domain Adversarial Neural Networks for Dysarthric Speech Recognition}
\name{Dominika Woszczyk$^1$, Stavros Petridis$^2$, David Millard$^3$}
\address{
  $^1$ $^2$ Department of Computing, Imperial College London, UK\\
  $^3$ School of Electronics and Computer Science, University of Southampton, UK}
\email{$^1$d.woszczyk19@imperial.ac.uk, $^2$stavros.petridis04@imperial.ac.uk, $^3$dem@ecs.soton.ac.uk}

\begin{document}
\bstctlcite{IEEEexample:BSTcontrol}

\maketitle
\begin{abstract}
Speech recognition systems have improved dramatically over the last few years, however, their performance is significantly degraded for the cases of accented or impaired speech. This work explores domain adversarial neural networks (DANN) for speaker-independent speech recognition on the UAS dataset of dysarthric speech. 
The classification task on 10 spoken digits is performed using an end-to-end CNN taking raw audio as input. The results are compared to a speaker-adaptive (SA) model as well as speaker-dependent (SD) and multi-task learning models (MTL). \textcolor{blue}{The experiments conducted in this paper show that DANN achieves an absolute recognition rate of 74.91\% and outperforms the baseline by 12.18\%. Additionally, the DANN model achieves comparable results to the SA model's recognition rate of 77.65\%. We also observe that when labelled dysarthric speech data is available DANN and MTL perform similarly, but when they are not DANN performs better than MTL.}

\end{abstract}
\noindent\textbf{Index Terms}: speech recognition, domain adversarial training neural networks, dysarthric speech, multi-task learning, transfer-learning

\section{Introduction}
The human voice is a powerful communication tool that makes it possible for us to interact with others but also with technology. Voice interfaces offer an easy and natural way of communication with devices. Unfortunately, speech recognition systems still perform relatively poorly on dysarthric speech. 

Dysarthria is a motor speech disorder that affects an individual's respiration, articulation, phonation and prosody, reducing their ability to produce intelligible speech. It is often the result of a neurological pathology  (e.g. Parkinson or Alzheimer) but it can also be the consequence of any brain-related injury. The impairment can take a multitude of patterns, each one depending on the severity and the cause of the disorder. However, this condition generally affects the tempo, rhythm or volume of an individual's speech. Due to the numerous causes of the disorder, dysarthria is classified into different types with distinguishable patterns, albeit each person has their differences. Nevertheless, the mistakes are consistent and predictable, while it could be hard to differentiate for a human ear, it makes it possible for a machine to adapt to an individual's speech.

With the advances of deep learning, automatic speech recognition (ASR) has reached high accuracy and robustness but still performs poorly on impaired speech. In a review of \cite{whistle}, the authors discuss the poor performance of ASR systems on dysarthric speech and call for necessary future work in this domain. In \cite{heysiri}, they show that current voice assistants reach on average an accuracy of 50-60\% on impaired speech while the minimal satisfactory rate for healthy speech is considered to be 90-95\% and is considered tolerable between 65\% to 80\% for individuals with speech impairment \cite{usability}.
The main issue is that deep learning architectures, with numerous parameters to optimise, need to train on a lot of data to achieve a good generalisation. Common ASR systems are trained on thousands of hours of healthy speech and often overlook speech with particularities such as accents or impairments.
Although many databases exist for dysarthric speech for multiple languages such as English (\cite{tapas, UAS,nemours}), Cantonese (\cite{can}) or Dutch (\cite{dysdutch}), they have a limited number of speakers and/or utterances. This is due to the difficulty of collecting data from individuals as they can be difficult to access and the effort required to produce large enough vocabulary is more exhausting than for healthy individuals. Looking into this problem, different domain adaptation architectures have been studied and successfully applied to dysarthric speech such as speaker-dependent (SD) and speaker-adaptive (SA) models. SA models showed particularly promising results \cite{ optimdys,spdspa,shor2019personalizing}. \textcolor{blue}{Nevertheless, those approaches still depend on available labelled data of the target speaker. Adversarial training using domain adversarial training of neural network (DANN) has been successfully applied to unsupervised domain adaptation in computer vision\cite{reversegd} and ASR\cite{datsp,rawdann}, making it an interesting approach to apply to dysarthric speech domain adaptation.}

This work investigates the DANN model on the task of 10 digits classification on the Universal Access Speech (UAS) dysarthric speech dataset. The model is compared to the performance of a multi-task learning model (MTL) as well as speaker-dependent and speaker-adaptive systems on an end-to-end baseline. 
The main contributions of the paper are as follows: 
\begin{itemize}
    \item \textcolor{blue}{To the authors' knowledge, this is the first work to apply end-to-end DANN architecture on the problem of dysarthric speech recognition and to compare it to other known solutions. We show that the model yields an overall word recognition rate (WRR) improvement of 12.18\% over the baseline.}
    \item \textcolor{blue}{We evaluate the results in the light of the satisfactory rate for dysarthric speech recognition (65\%) and observe that it is reached for 11 out of 15 speakers.} 
\end{itemize}

\section{Related work}
\subsection{Dysarthric speech}
Traditional machine learning is characterised by training and testing on data having the same distribution. However, collecting enough data is often either too hard or too expensive. This creates a difference of \textit{domains} between \textcolor{blue}{the data used}. In the case of an ASR system, the discrepancy can be caused by different pronunciations, and/or background noise. Multiple methods have been studied for dysarthric speech adaptation. One method is data augmentation with mono or multi-lingual healthy speech \cite{japadys} or simulated dysarthric speech by altering the tempo and the pitch \cite{dysaugm}. Alternatively, dysarthric speech can be turned into healthy speech using Generative Adversarial Networks (GANs) \cite{yang2020improving}. In \cite{dsy10}, Mel Frequency Cepstral Coefficients (MFCCs) are used with an artificial neural network (ANN) model to classify digits and silence from the UAS dataset. They achieve an overall WRR of 68.38\% for their speaker-independent model on three selected speakers (M07, M05, F05). In later work \cite{dsy25}, they extend the model to a multi-view multi-net ANN which is an ensemble model of several ANNs and reach an overall WRR of 85\%.
Other work also investigated the performance of speaker-adaptive models \cite{optimdys, spdspa}. Recently, the Euphonia project led by Google I/O collected through their platform a new database of non-standard speech to improve speech recognition for accented and/or impaired speech \cite{HowAIcan29:online}. In their paper \cite{shor2019personalizing}, they test sequence-to-sequence architectures, Bidirectional RNN Transducer (RNN-T) and Listen-Attention-Spell (LAS), trained on normal speech and fine-tuned on non-standard speech. They show that the adapted models perform better for both accented and dysarthric speech on all architectures. Additionally, the selection of the source data is important as it will affect the performance depending on the speakers' pronunciation patterns. The authors in \cite{selectinit}  propose a model that is selected based on the speaker severity level  which significantly improved the adaptation performance over the baseline model. \textcolor{blue}{In \cite{srcdomain20}, they used the entropy of posterior probability to select the data to train the model on before the adaptation step.}

\subsection{Domain adversarial training}
Domain adaptation models attempt to extract domain-invariant features that would generalise well on unseen domains, and minimise the distance between source and target domains.
There has been many studies on DNN domain adaptation such as work based on factorised hidden layer  \cite{lowres2018, sim2018domain},  i-vectors \cite{gupta2014vector}, KL-divergence regularisation \cite{kldiv2013}, or knowledge distillation models \cite{orbes2019knowledge,dnnkdis2017}.
Domain adversarial training neural networks (DANN), introduced in \cite{domain, reversegd}, is an unsupervised domain adaptation approach that tries to extract domain invariant features such that the model cannot determine whether the data belong to a domain or the other. They propose a reversal gradient layer (GRL) which introduces a negative gradient that maximises the loss of the classifier. The GRL can be easily implemented to any feed-forward network and without any additional parameter. This model provides the best improvements in supervised and semi-supervised setting and is similar to multi-task learning for supervised-learning. A multi-task learning (MTL) model is trained on multiple tasks simultaneously. With the introduction of an auxiliary task, the model is optimised to find parameters that perform well on more than one task, thus leading to better generalisation.
In \cite{multi}, the paper explores the multi-task model and accent embeddings to increase the robustness of ASR on unseen accents. In that model, the secondary task is an accent classifier while the main task is a multi accent acoustic model. They demonstrated that the MTL improves the performances over the single-stream network. 
Additionally, \cite{datsp} adapted the DANN model to accented speech and compared it to results given by an MTL architecture. They conclude that the DANN model outperforms the MTL model significantly when trained on unlabelled accented data.
We build upon previous work by exploring DANN and MTL for speaker-independent dysarthric speech recognition and compare it to previously established models and scenarios. Additionally, instead of using MFCCs or spectrograms as input features, the features from raw audio files are directly extracted by 1D-CNN layers as part of an end-to-end system.

\section{Methodology}
\subsection{Dataset}

The dysarthric speech was retrieved from the Universal Access Speech (UAS) database \cite{UAS}, due to the consistency, accessibility and the higher number of speakers than other datasets.
The dataset contains a total of 765 isolated words for each one of the 15 speakers (3 female and 12 male) with cerebral palsy and with different levels of severity: high, moderate and mild. Each speaker has either spastic\footnote{Stiffness of the muscles which makes movement difficult. The most common type of cerebral palsy.}, athetoid\footnote{Movement disorder caused by damage to the developing brain.} or mixed type of cerebral palsy.
Each word in the vocabulary is also repeated by 13 control speakers (4 female and 9 male). The collected data contains three batches of seven repetitions of digits, common words, 300 distinct uncommon words, computer commands and radio alphabet. \textcolor{blue}{For each speaker, their intelligibility level (0-100) has been computed by human listeners. To simplify the problem and training complexity, we retain the ten spoken digits only}.

\subsection{Models}
\subsubsection{Baseline model}
Convolutional neural networks (CNN) are often used to extract features and have been successfully applied in computer vision for pattern recognition, object detection or semantic segmentation. Additionally, CNN can also be applied to audio data by taking the signal in its raw form, or as a spectrogram or MFCCs. Work by \cite{raw1} and \cite{raw2} show that raw features can perform similarly to the pre-processed forms. RNN - based models have been successful on ASR tasks \cite{shor2019personalizing}. However, the disadvantage of recurrent architectures is that they are more complex to train and can overfit on smaller datasets.
We chose our baseline model to be a simpler 1D-CNN with seven layers and takes raw audio data as input, taking the whole word at once. The resulting features are fed to a fully connected layer with a sigmoid as an activation function. \textcolor{blue}{We selected the layers and parameters through experimentation (using grid search) on control and dysarthric speech independently, optimising for the WRR.}

\begin{figure}[!htb]
  \centering
  \includegraphics[width=8cm]{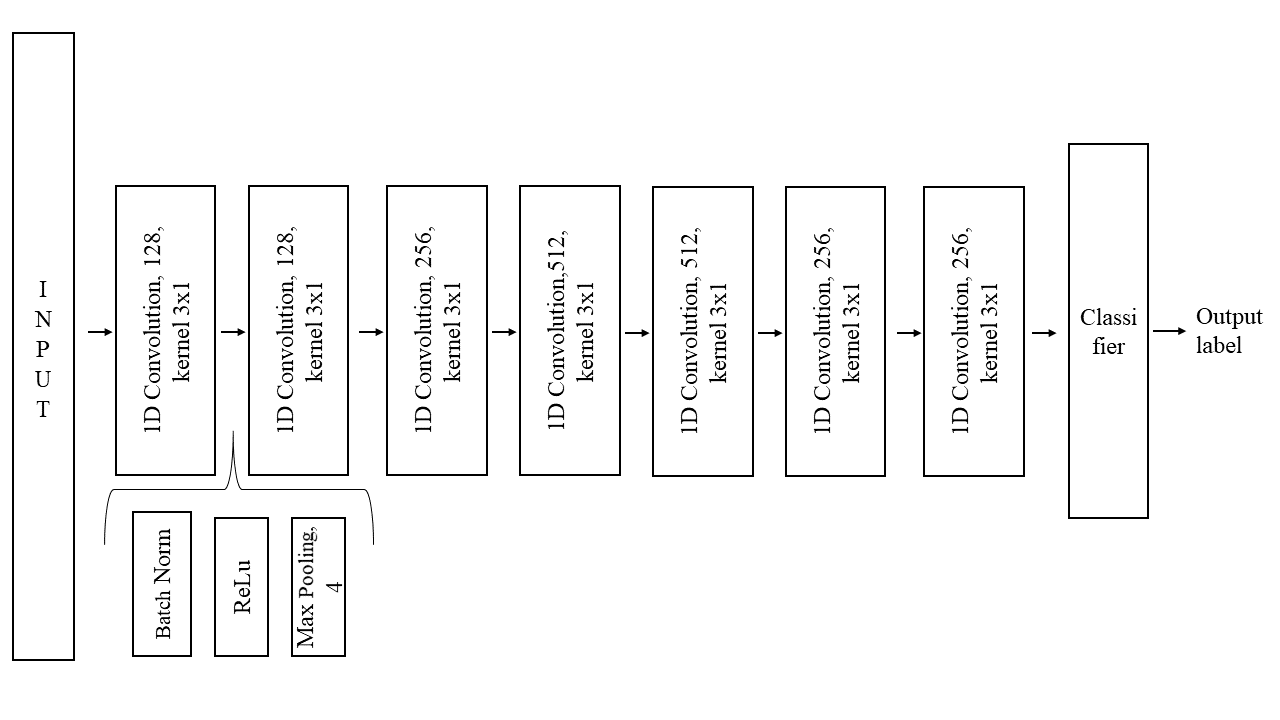}
  \caption{CNN baseline architecture}
  \label{Figure:cnn}
\end{figure}
 
\subsubsection{Domain adversarial training \& multi-task learning}

\begin{figure}[!htb]
  \centering
  \includegraphics[width=8.5cm]{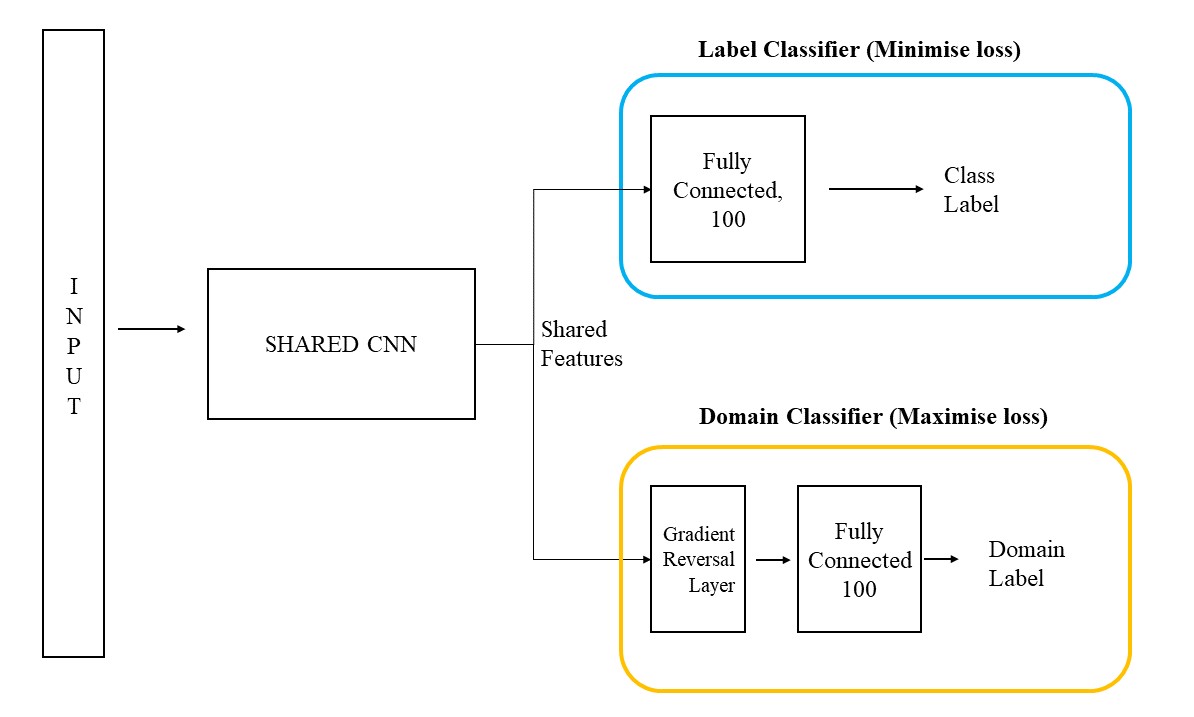}
  \caption{Domain adversarial neural network architecture. \textcolor{blue}{The label classifier outputs 0-9 digits. The domain classifier returns 0 for healthy speech and 1 for dysarthric speech.}}
  \label{Figure:dann}
\end{figure}

Following the work by \cite{reversegd}, we implement a gradient reversal layer (GRL) which outputs a negative gradient that allows the maximisation of the domain classification loss.
The DANN/MTL model we implemented has a shared block of seven convolutional layers just like the baseline as shown in Figure \ref{Figure:dann}. The model then branches out, one of the branches being the label classifier for the actual task trained on both the source and target domain. The second branch is the domain classifier with the GRL followed by a fully connected layer for the domain classification. 

The objective function of the task is given in \eqref{eq:loss}, \textcolor{blue}{where $N = n+n'$ is the total number of input samples} and $n$, $n'$ represent the control samples and dysarthric samples, $\theta_{f}$,$\theta_{d}$,$\theta_{y}$ are the parameters of the feature extractors, the domain classifier and the main task classifier. The parameters $y$ represents the classification label, $d$ domain labels and the indexes $s$, $t$ represent source (healthy) and target (dysarthric) domain respectively. The parameter $\alpha$ is a boolean such that for $\alpha =1$ we train the label classifier on target domain data as well.

\begin{equation}
  \footnotesize
\begin{array}{l}
E\left(\theta_{f}, \theta_{y}, \theta_{d}\right)=\frac{1}{n} \sum_{s=1}^{n} \mathcal{L}_{y}^{s}\left(\theta_{f}, \theta_{y}\right)
\\\quad +\alpha(\frac{1}{n^{\prime}} \sum_{s=n+1}^{N} \mathcal{L}_{y}^{t}\left(\theta_{f}, \theta_{y}\right)) 
\\
\quad-\lambda\left(\frac{1}{n} \sum_{s=1}^{n} \mathcal{L}_{d}^{s}\left(\theta_{f}, \theta_{d}\right)\right.
\left.+\frac{1}{n^{\prime}} \sum_{t=n+1}^{N} \mathcal{L}_{d}^{t}\left(\theta_{f}, \theta_{d}\right)\right)
\end{array}
\label{eq:loss}
\end{equation}

The MTL model follows the same architecture, with a normal layer instead of the GRL by setting $\lambda<0$.
Through experimentation, we set $\lambda$
to 1.5 and $-0.5$ for the DANN and MTL models.

Dysarthric speakers having their own particular speech patterns, 
each one of them could be a domain of their own. However, in this work, we consider all dysarthric speakers as one domain.

\subsection{Evaluation metric}
The word recognition rate (WRR) is used as the evaluation metric and is given by \eqref{eq:wrr} as 

\begin{equation}
  \footnotesize
WRR = \frac{\text{word correctly recognised}}{\text{words attempted}} *100
\label{eq:wrr}
\end{equation}

\section{Experiments \& results}


For all experiments, the same seed was used for the sake of reproducibility. The models were run for 30 epochs with early stopping performed using a hold-out validation set. We set the batch size to 32. 
\textcolor{blue}{For each speaker, the experiments were repeated three times rotating the batches of iterations and the average WRR was recorded.} 

\begin{figure}[!htb]
  \centering
  \includegraphics[width=6.5cm]{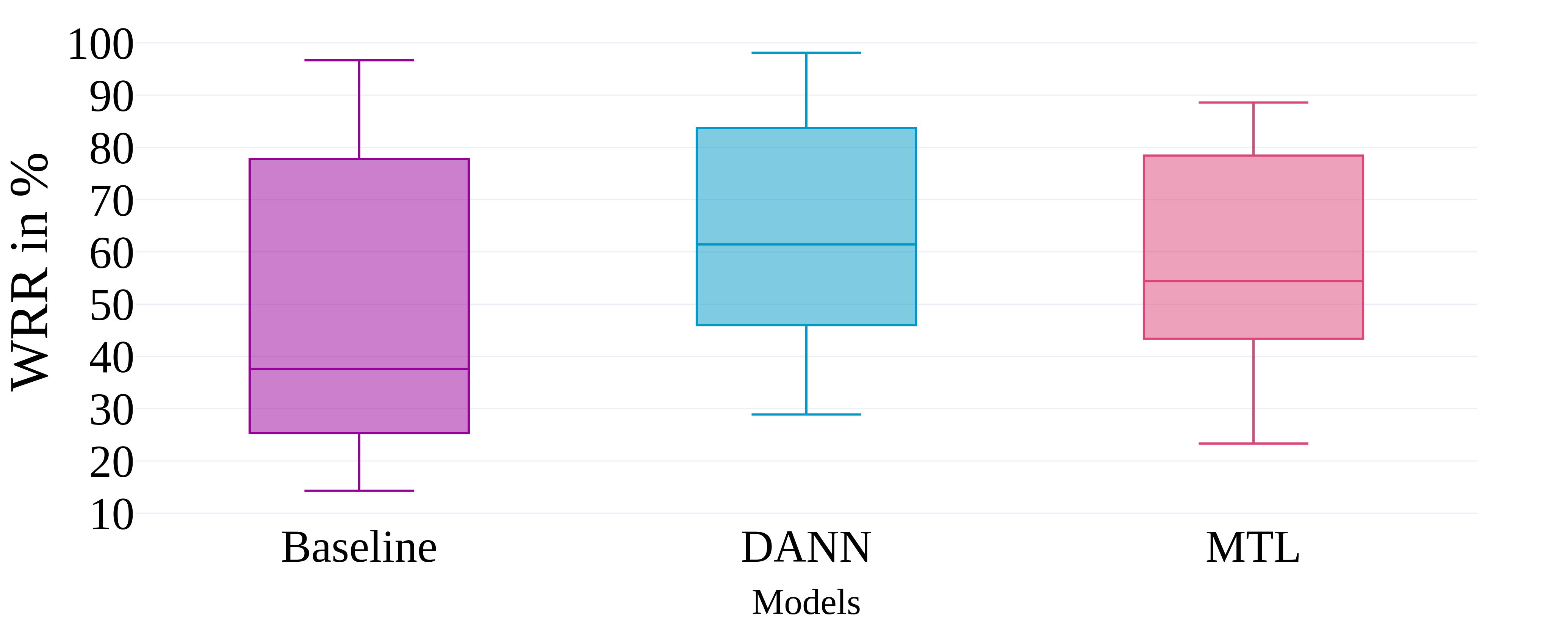}
  \caption{\textcolor{blue}{WRR distributions for the baseline, MTL and DANN on dysarthric speakers when trained with labelled control speech and unlabelled dysarthric speech.}}
  \label{Figure:wrr_unsup}
\end{figure}

\textcolor{blue}{\subsection{Training with unlabelled dysarthric speech}}
\textcolor{blue}{In the first experiment, we want to see how DANN, MTL and the baseline perform when we do not have any labelled data for dysarthric speech available}. We train the baseline on the control speech only (2730 samples) and test on all the repetitions for each dysarthric speaker (210 samples per speaker). For the MTL and DANN models, we set $\alpha$ = 0 and \textcolor{blue}{we train them on control speech and all dysarthric speakers' unlabelled iterations excluding the test subject}, for which the first batch of iterations (70 samples) was used as a validation set and the two other batches were used in the test set (140 samples). The results on Table \ref{Table:unsupervised} show that the DANN model performs better than the baseline and the MTL model on average for all groups. One can see on Figure~\ref{Figure:wrr_unsup} that both DANN and MTL have a lesser variability than the baseline as well as DANN have a higher median than both the baseline and the MTL model.



\begin{table}[!htb]
\caption{\textcolor{blue}{Recognition rates for the baseline, MTL and DANN on dysarthric speakers when trained with labeled control speech and unlabelled dysarthric speech. The number in bold is the best WRR for its row.}}
\centering
\scriptsize
\begin{tabular}{lllll}
\hline \toprule
Severity & Speaker & Baseline  & DANN   & MTL   \\ \hline
Mild     & F05     & 87.62 & 85.71  & \textbf{88.10}  \\
         & M08     & 96.67 & \textbf{97.14}  & 80.48 \\
         & M10     & 96.19 & \textbf{98.10}   & 88.57 \\
         & M14     & \textbf{92.86} & 88.10   & 88.57 \\
         & M09     & 25.71 & \textbf{75.24}  & 55.24 \\ \hline
Moderate      & M11     & 29.44 & 53.89  & \textbf{54.44} \\
         & F04     & 38.10  & \textbf{56.67}  & 51.90 \\
         & M05     & 37.62 & \textbf{77.62}  & 61.90  \\ \hline
High     & M16     & 48.33 & \textbf{76.67}  & 72.22 \\
         & F02     & 40.48 & \textbf{45.24}  & 41.90  \\
         & M07     & 25.24 & \textbf{61.43}  & 46.67 \\
         & M01     & 22.73 & 48.18  & \textbf{51.82} \\
         & M12     & 21.11 & \textbf{28.89}  & 23.33 \\
         & F03     & 14.29 & 31.90  & \textbf{34.76} \\
         & M04     & 29.53 & 39.60   & \textbf{42.28} \\ \hline
Mean (SD) & Mild  & 79.81 (30.46) & \textbf{88.86 (9.35)}  & 80.19 (14.37)\\
         & Mod     & 35.05 (4.87) & \textbf{62.73 (12.97)}  & 56.08 (5.20) \\
          & High    & 28.82 (11.82) & \textbf{47.42 (16.85)}  & 44.71 (15.17) \\
         & All    & 47.06 (30.14) & \textbf{64.29 (23.00)} & 58.81 (20.63) \\ \bottomrule
\end{tabular}
 \label{Table:unsupervised}
\end{table}

\textcolor{blue}{\subsection{Training with labelled  dysarthric speech}}  
The second set of experiments are conducted to compare the performance of the DANN, MTL and the baseline model as a speaker-independent (SI) model (none of the utterances of the speaker was present in the training set) when some labelled dysarthric speech is available. In this case, we set $\alpha=1$ for both DANN and MTL.
The baseline model was trained on all control speakers (2730 samples) and the first batch of iterations of the tested dysarthric speaker was used as a validation (70 samples) set and the two other batches were used in the test set (140 samples).

Additionally, we compare the models to the baseline when it is trained as a speaker-dependent (SD) and speaker-adaptive (SA) system.
\textcolor{blue}{The SA model is the SI model fine-tuned with one of the three iterations of the tested speaker and validated and tested with the two remaining iterations.} Through experimentation, fine-tuning only the last convolutional layer and the linear classifier yields the best results. Therefore those are the ones shown in Table \ref{Table:res}.
The SD model was trained on control speakers combined with \textcolor{blue}{one iteration of the tested subject (2800 samples), the other two iterations were used for validation (70 samples) and testing (70 samples).}

\begin{table}[!htb]
\caption{\textcolor{blue}{Recognition rates for models trained on labelled dysarthric and control speech. The number in bold is the best WRR for its row.}}
\centering
\resizebox{\columnwidth}{!}
{
\begin{tabular}{lllllll}\toprule
       Severity  & Speaker    & SI    & DANN  & MTL   & SD    & SA    \\ \midrule
Mild     & F05 & 86.67 & 90.48 & 90.48 & 89.21 & \textbf{92.86} \\
         & M08 & 95.24 & 95.24 & \textbf{98.10}  & 96.19 & 94.29 \\
         & M10 & \textbf{100.00}   & 98.58 & 98.00    & 98.86 & \textbf{100.00}   \\
         & M14 & \textbf{100.00}   & 94.29 & 94.00    & 96.10 & \textbf{100.00}   \\
         & M09 & 53.57 & 92.14 & 90.00    & 78.57 & \textbf{94.29} \\ \midrule
Moderate & M11 & 50.00    & \textbf{88.89} & 85.56 & 74.82 & 85.56    \\
         & F04 & 67.14 & 75.71 & \textbf{77.81} & 73.55 & 72.86 \\
         & M05 & 70.48 & 86.19 & 89.52 & 82.06 & \textbf{100.00}   \\ \midrule
High     & M16 & 65.56 & 76.67 & \textbf{82.22} & 74.82 & 80.00    \\
         & F02 & 53.33 & 48.10  & 45.71 & 49.05 & \textbf{72.86} \\
         & M07 & 45.24 & 75.24 & 72.38 & 64.29 & \textbf{84.29} \\
         & M01 & 41.82 & 62.73 & \textbf{65.45} & 56.67 & 53.33 \\
         & M12 & 30.56 & 25.00    & 26.11 & 27.22 & \textbf{45.00}    \\
         & F03 & 50.48 & 70.00    & \textbf{71.43} & 63.97 & 50.00    \\
         & M04 & 30.87 & 44.30  & \textbf{44.97} & 40.05 & 40.00 \\ \hline
Mean (SD) & &&&&&\\
&Mild & 87.10 (23.11) & 94.15 (3.50) & 94.12 (4.94) & 91.79 (10.09) &\textbf{96.29 (5.35)} \\
         & Mod & 62.54 (10.99) & 83.60 (6.96) & 84.30 (5.96) & 76.81 (4.59) & \textbf{85.95 (13.58)} \\
          & High & 45.41 (12.51) & 57.43 (19.08) & 58.32 (19.86) & 53.72 (16.24) & \textbf{60.78 (17.89)}\\
         & All &62.73 (23.56)    &74.91 (21.60)    &75.45 (21.59)    &71.03 (21.12) &\textbf{77.65 (21.18)}\\
\bottomrule
\end{tabular}}
 \label{Table:res}
\end{table}

\begin{figure}[!htb]
  \centering
  \includegraphics[width=6.5cm]{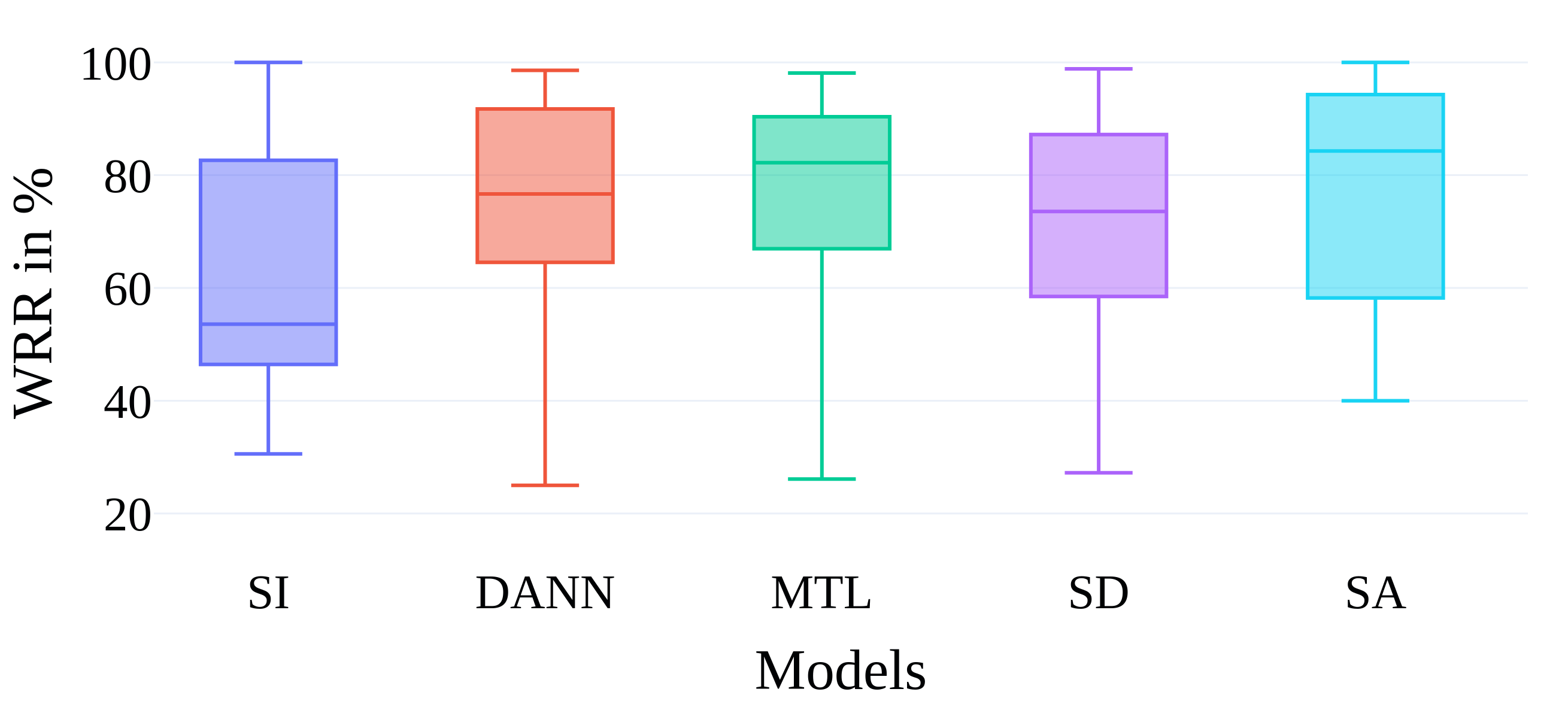}
  \caption{WRR distributions for all severity subjects for models trained with labelled control and dysarthric speech.}
  \label{Figure:wrr_all}
\end{figure}

The results on Table \ref{Table:res} show that the baseline SI model reached an overall WRR of (62.73 $\pm$23.56)\% and is the worst-performing model for almost all speakers, with some exceptions. Due to the small number of training samples, the model overfits relatively fast. Each dysarthric speaker has a unique impairment which makes it harder for a model to generalise well on multiple speakers when it has not been introduced to the test subject specifically. However, this model achieved similar mean WRR of 67.46\% for the M05, F05 and M07 speakers as the SI model from \cite{dsy10} (68.38\%), which used carefully selected MFCCs. 

On the other hand, the best performing system is the SA model, reaching the highest WRR reaching an average of (77.65 $\pm$21.18)\%, and severity specific means of 96.28\%, 85.96\% and 60.78\% for mild, moderate and high levels. This confirms previous work \cite{optimdys,spdspa,shor2019personalizing} that showed that by fine-tuning on a target subject it is possible to improve greatly the performance of the ASR system. However, Figure \ref{Figure:wrr_all} shows that there is a high variability in the performance comparable to the SI model and greater than for the other models. 
Table \ref{Table:res} also shows that both DANN and MTL models outperform the baseline model for both SD and SI scenarios. Moreover, both models reach similar performance to the speaker adaptive system for all severity levels, outperforming it at times. However, the DANN and MTL models perform similarly as shown in Figure \ref{Figure:wrr_all} but the MTL model has a higher median and a lower variance indicating a more consistent performance.

Although MTL has a slightly better average WRR (0.54\% difference), none of the models is significantly better than the other. Both models introduce regularisation by extraction domain invariant features, with the difference being that the DANN model attempts to confuse the domain classifier and maximise the loss while MTL is minimising its loss. In this scenario, MTL achieves a slightly higher mean.
In terms of the SD model which achieved on average a WRR of (71.03 $\pm$21.12)\%, it performs better than the SI one but is outperformed by the SA model. Because of the small training set, the SD system was trained on, it is easier to overfit. However, the SA model benefits from the pre-training on the bigger dataset. Similarly, the SD model is outperformed by the MTL and DANN models. Those models also benefit from more generalised features as they were trained in a multi-task manner and on a bigger dataset as well. 

Regarding usability, from results on Table \ref{Table:res} we can see that the SA, DANN and MTL models achieve satisfactory recognition rates at healthy speech level (90\%) for all speakers with mild severity dysarthria. The systems also reach the tolerable level for dysarthric users for all subjects at moderate severity with no WRR below 70\%. For high severity speakers, all reach three out of seven. In total, all three best-performing algorithms achieve a satisfactory WRR for 11 out of 15 speakers.

\section{Conclusion}

Speech recognition has made great progress with the advances of deep learning architectures. However, it still lacks robustness to non-standards speech for which there is usually less labelled data available. This paper leverages domain-invariant features with domain adversarial training to cope with limited data on the UAS dataset of dysarthric speech. In the conducted experiments, domain adversarial training neural networks (DANN) and multi-task-learning (MTL) models were studied for cases when some labelled dysarthric data is and is not available. We compare them to a baseline CNN model trained on speaker-dependent (SD) and speaker-independent (SI) scenarios and a speaker-adaptive (SA) system by fine-tuning each model on each speaker. Results show that the DANN architecture can significantly improve over the baseline CNN for speakers of all levels of severity when some labelled data is available. Moreover, the results are similar to the best performing model fine-tuned on each speaker (SA), outperforming it for some subjects. We also observe that in this supervised scenario MTL and DANN perform very similarly but the DANN model outperforms the MTL model when trained with no target domain data. Furthermore, our end-to-end baseline achieves similar WRR to a model from previous work using MFCCs. Finally, we discuss the usability of the models and observe that the DANN model achieve a satisfactory WRR for dysarthric speakers for 11 out of 15 subjects.

\bibliographystyle{IEEEtran}

\bibliography{mybib}

\end{document}